\documentstyle[12pt,epsf]{article}
\begin{document}
 
\thispagestyle{empty}
\begin{flushright}
CERN-TH/2001-292\\
October 2001 
\end{flushright}

\vspace*{1.5cm}
\centerline{\Large\bf Rare Kaon Decays -- Overview}
\vspace*{2cm}
\centerline{{\sc Gerhard Buchalla}}
\bigskip
\centerline{\sl Theory Division, CERN, CH-1211 Geneva 23,
                Switzerland}
 
\vspace*{1.5cm}
\centerline{\bf Abstract}
\vspace*{0.3cm}
\noindent 
The theory of rare $K$ decays is reviewed, emphasizing
short-distance processes and the prospects to probe the
physics of flavour. A brief overview of the subject
is presented, along with a more detailed discussion
of the theory of $K\to\pi\nu\bar\nu$ decays.

\vspace*{5cm}
\centerline{\it Invited Talk presented at KAON 2001}
\centerline{\it International Conference on CP Violation,
                Pisa, 12-17 June 2001}
\vfill

\begin{flushleft}
CERN-TH/2001-292
\end{flushleft}
 
\newpage
\pagenumbering{arabic}

\section{Introduction and Overview}

Rare decays of kaons \cite{REV} probe the details of weak interactions
at the quantum level. They can be sensitive to energy scales
much higher than the kaon mass itself and can thus yield
fundamental insights into physics at very short distances.
A remarkable historical example is the suppression of
flavour-changing neutral currents, implied by the fact that
$B(K_L\to\mu^+\mu^-)=7\cdot 10^{-9}$ while
$B(K^+\to\mu^+\nu)=0.64$, which led to the GIM mechanism and the
concept of charm. Another famous case is the $K_L$--$K_S$ mass
difference $\Delta M_K$, which arises through $K$--$\bar K$ mixing,
a second-order weak process with high sensitivity to the
charm-quark mass $m_c$. The analysis of this rare transition by
Gaillard and Lee in 1974 gave a correct estimate of $m_c\sim 1.5$ GeV,
prior to the discovery of charm.
Today the focus has shifted to modes that depend on much
higher energy scales, related to CP violation, the top quark
and, in general, new degrees of freedom from physics beyond
the standard model. However, the spirit of the approach is still
very much the same.

The field of rare $K$ decays is rich and varied.
Three broad classes may be distinguished:

\begin{itemize}
\item
Long-distance dominated rare or radiative decays such as
$K^+\to\pi^+l^+l^-$, $K_L\to\pi^0\gamma\gamma$, $K_S\to\gamma\gamma$
or $K_L\to\mu^+\mu^-$. Although not immediately useful to obtain
short-distance information, they are still important to learn about
low-energy QCD dynamics, largely relying on chiral perturbation
theory. In this way long-distance ``background'' can be better controlled
in cases where it is more difficult to disentangle short- and long-distance
dynamics.
\item
Short-distance dominated decays as $K^+\to\pi^+\nu\bar\nu$ and
$K_L\to\pi^0\nu\bar\nu$ can provide excellent tools to test the standard
model with high precision. $K_L\to\pi^0e^+e^-$ partly belongs to
this class as well, but long-distance physics plays a non-negligible
role in this case.
\item
Decay modes that are forbidden in the standard model could be
dramatic indicators of new physics. Examples are
$K_L\to\mu e$, $K^+\to\pi^+\mu^+ e^-$ and $K_L\to\pi^0\mu e$,
where stringent upper limits on the branching ratios exist of 
$4.7\cdot 10^{-12}$ \cite{BNL871}, 
$2.8\cdot 10^{-11}$ \cite{BNL865} and 
$4.4\cdot 10^{-10}$ \cite{KTeV:lfv}, respectively.
\end{itemize}

In the following section we will very briefly discuss the most
important theoretical methods needed to describe rare $K$ decays.
Subsequently we shall focus on the ``golden modes'' 
$K^+\to\pi^+\nu\bar\nu$ and $K_L\to\pi^0\nu\bar\nu$. We
review in particular the status of theoretical uncertainties,
especially in the charm sector of $K^+\to\pi^+\nu\bar\nu$,
which is the most critical issue.
For a more detailed account of other important subjects in the field
of rare kaon decays we refer the reader to the corresponding
articles in these proceedings. Specific overviews are given
by D'Ambrosio (long-distance modes), Silvestrini (new physics)
and Littenberg (experiment).

\section{Theoretical Methods}

\begin{figure}[t]
  \vspace{5.0cm}
  \includegraphics{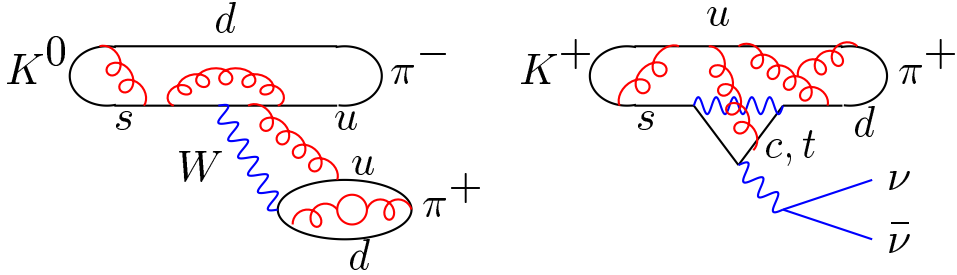}
  \caption{\it QCD effects in weak decays.
    \label{fig:kppkpn} }
\end{figure}

The essential problem in computing weak decays of hadrons
is the influence of strong interactions, which need to be
properly accounted for to extract the underlying flavour
physics at the quark level (Fig. \ref{fig:kppkpn}).
For kaon decays two different, complementary approaches are at
our disposal: The framework of effective weak hamiltonians
and chiral perturbation theory (ChPT).

The {\it effective weak hamiltonian\/} has the form
\begin{equation}\label{heffw}
{\cal H}_{eff}=\frac{G_F}{\sqrt{2}} V_{CKM}
\sum_i\, C_i(\mu) Q_i
\end{equation}
Here $V_{CKM}$ is a CKM factor. The $C_i$ are Wilson coefficients
comprising the short-distance contributions from scales
$>\mu\sim 1$ GeV. They can be calculated perturbatively from the
underlying fundamental theory, i.e. the standard model or one of
its extensions. The $Q_i$ are dimension-6 operators, typically of
the 4-fermion type such as e.g. $(\bar su)_{V-A}(\bar ud)_{V-A}$.
Their matrix elements between hadronic states contain the
nonperturbative, long-distance dynamics of the weak amplitude.
\begin{itemize}
\item
The effective hamiltonian formalism is based on an operator product
expansion (corresponding to integrating out heavy fields such
as $W$ or top), which achieves a systematic factorization of
short-distance ($C_i$) and long-distance ($\langle Q_i\rangle$)
contributions.
\item
The disadvantage is that using ${\cal H}_{eff}$ the QCD dynamics
is still formulated in terms of quarks and gluons. The matrix
elements of the $Q_i$, involving quarks, must be calculated between
hadronic states. In general, this is a very complicated task.
\item
The advantage is that the short-distance information
(dependence on top, weak phases, new physics parameters), which
we are primarily interested in, is explicit.
\end{itemize}

In {\it chiral perturbation theory\/} the strong and weak interactions
are formulated in terms of the meson fields
\begin{equation}\label{sigphi}
\Sigma =\exp\left(\frac{2i}{f}\Phi\right)
\end{equation}
where 
\begin{equation}\label{phi33}
\Phi\equiv T^a\pi^a=
\left(\begin{array}{ccc}
\frac{\pi^0}{\sqrt{2}}+\frac{\eta}{\sqrt{6}} & \pi^+ & K^+ \\
\pi^- & -\frac{\pi^0}{\sqrt{2}}+\frac{\eta}{\sqrt{6}} & K^0 \\
K^- & \bar K^0 & -\frac{2\eta}{\sqrt{6}}
\end{array}\right)
\end{equation}
$\Sigma$ transforms as $\Sigma \to L\Sigma R^\dagger$ under
$SU(3)_L\otimes SU(3)_R$ chiral rotations and 
$f$ is the generic decay constant
for the light pseudoscalars (in a normalization in which
$f_\pi=131\,{\rm MeV}$).
The lagrangians of strong ($QCD$) and weak ($\Delta S=1$)
interactions are organized in powers of derivatives,
corresponding to momenta, and quark masses:
\begin{equation}\label{lqcdchi}
{\cal L}^{QCD}={\cal L}^{QCD}_2 +  {\cal L}^{QCD}_4 +\ldots
\end{equation}
\begin{equation}\label{lds1chi}
{\cal L}^{\Delta S=1}=
{\cal L}^{\Delta S=1}_2 +  {\cal L}^{\Delta S=1}_4 +\ldots
\end{equation}
The lowest order terms read explicitly
\begin{equation}\label{lqcd2}
{\cal L}^{QCD}_2=\frac{f^2}{8}\,{\rm tr}
\left[ D_\mu\Sigma D^\mu\Sigma^\dagger +
2 B_0({\cal M}\Sigma^\dagger +\Sigma {\cal M}^\dagger)\right]
\end{equation}
\begin{equation}\label{lds12}
{\cal L}^{\Delta S=1}_2=\frac{G_F}{\sqrt{2}}|V^*_{us}V_{ud}|g_8
\frac{f^4}{4}\, {\rm tr}\, \lambda_6 D_\mu\Sigma D^\mu\Sigma^\dagger
\end{equation}
Here ${\cal M}={\rm diag}(m_u,m_d,m_s)$ is the quark mass matrix,
$f$ and $B_0$ are the two parameters entering ${\cal L}^{QCD}_2$,
and $g_8$ is the dominant weak coupling in ChPT at second
order in momenta (for simplicity we have omitted a second term
in ${\cal L}^{\Delta S=1}_2$, proportional to a coupling
$g_{27}$, which is suppressed numerically as a consequence of
the $\Delta I=1/2$ rule).
\begin{itemize}
\item
ChPT is a low-energy effective theory based on the chiral
symmetries of QCD combined with an expansion in $p^2/\Lambda^2$,
where $p$ represents the (small) momenta of the light mesons
and $\Lambda\sim 1$ GeV is the hadronic scale.
This approach is model-independent, since all possible interaction
terms with the correct properties under chiral symmetry have
to be included. Each term is multiplied with an a priori unknown
parameter (or coupling constant). The framework becomes predictive
because only a finite number of parameters appears to any fixed order
in the momentum expansion. Once they are extracted from a corresponding
number of measurements, further predictions can be made.
\item
The advantage of ChPT is that the QCD dynamics is already
expressed in terms of hadrons ($\Phi$).
\item
The disadvantage is that the short-distance physics is
implicit. It is hidden in the counterterms, i.e. the
coupling constants of the chiral lagrangian, which are renormalized
by loop contributions.
\end{itemize}
In principle the ``dual'' pictures of ${\cal L}^{\Delta S=1}$
and ${\cal H}_{eff}$ describe the same physics. However, 
establishing a direct link between them would require the
computation of the matrix elements $\langle Q_i\rangle$ and
a comparison of the resulting $\langle{\cal H}_{eff}\rangle$
with the amplitudes from ${\cal L}^{\Delta S=1}$.
For a general kaon decay this is not possible at present because
of our poor control of nonperturbative QCD in terms of quarks
and gluons. In the meantime, the two pictures approach the problem
of weak amplitudes from opposite directions:
${\cal H}_{eff}$ starting from high energies, 
${\cal L}^{\Delta S=1}$ from low energies.
From the characteristic advantages and disadvantages listed
above it is clear that ${\cal H}_{eff}$ is more useful for
applications where short-distance dynamics is essential,
such as short-distance dominated rare decays or CP violation
($\varepsilon$, $\varepsilon'/\varepsilon$). On the other hand,
ChPT is the method of choice for processes controlled by
long-distance physics.

An important special case is given by the modes $K\to\pi\nu\bar\nu$.
Here the hadronic matrix element of the quark-level operator
is particularly simple and known from $K\to\pi l\nu$.
In this situation the effective hamiltonian approach can solve
the problem completely as we will further discuss in the
following section.
It is interesting to note that a complementary example exists as well:
$K_S\to\gamma\gamma$ can be computed by a finite one-loop
calculation based on (\ref{lds12}). In that case a parameter-free
prediction is obtained once $g_8$ is fixed from $K\to\pi\pi$
(see the talk by D'Ambrosio for more details).

\section{The Golden Modes:\\
$K^+\to\pi^+\nu\bar\nu$ and $K_L\to\pi^0\nu\bar\nu$}

\subsection{Basic Properties and Results}

\begin{figure}[t]
  \vspace{4.0cm}
  \includegraphics{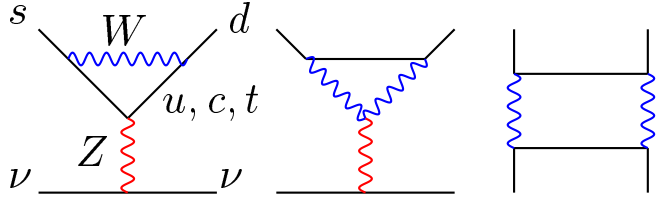}
  \caption{\it The leading order electroweak diagrams contributing to
               $K\to\pi\nu\bar\nu$ in the standard model.
    \label{fig:kpnn} }
\end{figure}

The decays $K\to\pi\nu\bar\nu$ proceed through flavour-changing
neutral currents, which arise at one loop in the standard model
(Fig. \ref{fig:kpnn}).
The GIM structure of the amplitude can be written as
\begin{equation}\label{gimf}
\sum_{i=u,c,t}\lambda_i\, F(x_i)=
\lambda_c\, (F(x_c)-F(x_u))+\lambda_t\, (F(x_t)-F(x_u))
\end{equation}
with $\lambda_i=V^*_{is}V_{id}$ and $x_i=m^2_i/M^2_W$.
The first important point is the characteristic {\it hard\/}
GIM cancellation pattern, which means that the function $F$
depends as a power on the internal mass scale
\begin{equation}\label{fuct}
F(x_u)\sim\frac{\Lambda^2_{QCD}}{M^2_W}\sim 10^{-5}
\ll F(x_c)\sim \frac{m^2_c}{M^2_W}\ln\frac{M_W}{m_c}\sim 10^{-3}
\ll F(x_t)\sim 1
\end{equation}
The up-quark contribution is a long-distance effect,
determined by the scale $\Lambda_{QCD}$.
As an immediate consequence, top and charm contribution
with their hard scales $m_t$, $m_c$
dominate the amplitude, whereas the long-distance part $F(x_u)$
is negligible. Note that the charm contribution,
$\lambda_c\, F(x_c)\sim 10^{-1}\cdot 10^{-3}$, and the top
contribution,  $\lambda_t\, F(x_t)\sim 10^{-4}\cdot 1$, have the
same order of magnitude when the CKM factors are included.
The short-distance dominance of the $s\to d\nu\bar\nu$ transition
next implies that the process is effectively semileptonic, because
a single, local operator $(\bar sd)_{V-A}(\bar\nu\nu)_{V-A}$
describes the interaction at low-energy scales.
Hence the amplitude has the form
\begin{equation}\label{akpnn}
A(K^+\to\pi^+\nu\bar\nu)\sim
G_F\alpha(\lambda_c F_c+\lambda_t F_t)
\langle\pi^+|(\bar sd)_V|K^+\rangle\, (\bar\nu\nu)_{V-A}
\end{equation}
The coefficient function $\lambda_c F_c+\lambda_t F_t$ is
calculable in perturbation theory. The hadronic matrix element
can be extracted from $K^+\to\pi^0 e^+\nu$ decay via isospin.
The $K^+\to\pi^+\nu\bar\nu$ amplitude is then completely determined,
and with good accuracy.

The neutral mode proceeds through CP violation in the standard model
and has the form
\begin{equation}\label{aklpn}
A(K_L\to\pi^0\nu\bar\nu)\sim 
{\rm Im}\lambda_t\, F_t + {\rm Im}\lambda_c\, F_c
\end{equation}
where
\begin{equation}\label{imftc}
{\rm Im}\lambda_t\, F_t \sim 10^{-4}\cdot 1
\gg {\rm Im}\lambda_c\, F_c \sim 10^{-4}\cdot 10^{-3}
\end{equation}

The $K\to\pi\nu\bar\nu$ modes have been studied in great detail
over the years to quantify the degree of theoretical precision.
Important effects come from short-distance QCD corrections.
These were computed at leading order in \cite{DDG}.
The complete next-to-leading order  calculations \cite{BB123,MU,BB99}
reduce the theoretical uncertainty in these decays to
$\sim 5\%$ for $K^+\to\pi^+\nu\bar\nu$ and $\sim 1\%$ for
$K_L\to\pi^0\nu\bar\nu$.
This picture is essentially unchanged when further small effects
are considered, including isospin breaking in the relation of
$K\to\pi\nu\bar\nu$ to $K^+\to\pi^0l^+\nu$ \cite{MP},
long-distance contributions
\cite{RS,HLLW}, the CP-conserving effect in $K_L\to\pi^0\nu\bar\nu$
in the standard model \cite{RS,BI}, two-loop electroweak 
corrections for large $m_t$ \cite{BB7} and subleading-power
corrections in the OPE in the charm sector \cite{FLP}.

While already $K^+\to\pi^+\nu\bar\nu$ can be reliably calculated,
the situation is even better for $K_L\to\pi^0\nu\bar\nu$. Since
only the imaginary part of the amplitude 
contributes, the charm sector, in $K^+\to\pi^+\nu\bar\nu$
the dominant source of uncertainty, is completely negligible for
$K_L\to\pi^0\nu\bar\nu$ ($0.1\%$ effect on the branching ratio).
Long distance contributions 
($\;\raisebox{-.4ex}{\rlap{$\sim$}} \raisebox{.4ex}{$<$}\; 0.1\%$)  
and also the indirect CP violation effect  
($\;\raisebox{-.4ex}{\rlap{$\sim$}} \raisebox{.4ex}{$<$}\; 1\%$)  
are likewise negligible. 
The total theoretical
uncertainties, from perturbation theory in the top sector
and in the isospin breaking corrections, are safely below
$3\%$ for $B(K_L\to\pi^0\nu\bar\nu)$. This makes this decay
mode truly unique and very promising for phenomenological
applications.

In Table \ref{kpnntab} we have summarized some of the main
features of $K^+\to\pi^+\nu\bar\nu$ and $K_L\to\pi^0\nu\bar\nu$.
\begin{table}
\centering
\caption{\it Compilation of important properties and results
for $K\to\pi\nu\bar\nu$.
}
\vskip 0.1 in
\begin{tabular}{|c|c|c|} \hline
& $K^+\to\pi^+\nu\bar\nu$ & $K_L\to\pi^0\nu\bar\nu$ \\
\hline
\hline
& CP conserving & CP violating \\
\hline
CKM & $V_{td}$ & $\mbox{Im} V^*_{ts}V_{td}\sim J_{CP}\sim\eta$ \\
\hline
contributions & top and charm & only top \\
\hline
scale dep. (BR) &  $\pm 20\%$ (LO) &
                   $\pm 10\%$ (LO) \\
                &  $\to\pm 5\%$ (NLO) &
                   $\to\pm 1\%$ (NLO) \\
\hline
BR (SM) & $(0.8\pm 0.3)\cdot 10^{-10}$&$(2.6\pm 1.2)\cdot 10^{-11}$ \\
\hline
exp.  & $\left(1.5^{+3.4}_{-1.2}\right)\cdot 10^{-10}$ BNL 787 \cite{SAT}
           & $< 5.9\cdot 10^{-7}$ KTeV \cite{KLPNTEV} \\
\hline
\end{tabular}
\label{kpnntab}
\end{table}
Note that the ranges given as the standard model
predictions in Table \ref{kpnntab} arise from our, at present,
limited knowledge of standard model parameters (CKM), and not
from intrinsic uncertainties in calculating 
the branching ratios.


\subsection{Theoretical Uncertainties in $K^+\to\pi^+\nu\bar\nu$
and $|V_{td}|$}

Table \ref{tab:vtdkpn}
\begin{table}[t]
  \centering
  \caption{\it Relative uncertainties in $|V_{td}|$ from
   $K^+\to\pi^+\nu\bar\nu$. The errors shown added in quadrature
   amount to a total of $\Delta |V_{td}|/|V_{td}|={\pm 12\%}$.}
  \vskip 0.1 in
\begin{tabular}{|c|c|c|c|}
\hline
\multicolumn{2}{|c|}{{$B(K^+\to\pi^+\nu\bar\nu)$}} & 
        {$\mu_t/$GeV} & {$\mu_c/$GeV} \\
\hline
\multicolumn{2}{|c|}{{$(1.0\pm 0.1)\cdot 10^{-10}$}} & 
        {100 -- 300} & {1 -- 3}  \\
\hline
\multicolumn{2}{|c|}{{$\pm 6.8\%$}} & {$\pm 0.5\%$} &
      {$\pm 4.5\%$}  \\
\hline
\hline
$V_{cb}$ & $\Lambda^{(4)}_{\overline{\rm MS}}/$GeV & $m_t/$GeV & $m_c/$GeV \\
\hline
$0.040\pm 0.002$ & $0.325\pm 0.080$ & $166\pm 5$ & $1.3\pm 0.1$ \\
\hline
{$\pm 5.1\%$} & {$\pm 2.2\%$} & {$\pm 3.5\%$} & 
           {$\pm 5.8\%$} \\
\hline
\end{tabular}
  \label{tab:vtdkpn}
\end{table}
displays the various sources of uncertainty for determining
$|V_{td}|$ from $K^+\to\pi^+\nu\bar\nu$, where a measured
branching ratio of $(1.0\pm 0.1)\cdot 10^{-10}$ is assumed for
illustration. Uncertainties from input parameters are shown in
the lower half. The residual dependences on the renormalization
scales ($\mu_t$ and $\mu_c$ for the top-sector and the charm-sector,
respectively) are used to estimate the uncertainty intrinsic to
the theoretical calculation itself, which is necessarily
approximate and relies here on NLO perturbation theory.

The most critical issue is clearly the charm sector. After all,
$m_c\equiv\bar m_c(\bar m_c)=1.3\,{\rm GeV}$ is not extremely large
compared to $\Lambda_{QCD}$ and the applicability of the OPE and
perturbation theory at the charm scale has to be decided on a 
case-by-case basis.
Concerning the reliability of perturbation theory in the present
case, the following checks can be made.
\begin{itemize}
\item
The $\mu_c$-dependence is reduced from $\pm 28\%$ at LO down
to  $\pm 13\%$ at NLO in RG improved perturbation theory.
(Here and in the following discussion these are relative uncertainties
refering to the charm amplitude alone. Due to the existence
of the large top-quark contribution, their impact is reduced
in the $|V_{td}|$ determination. For instance, the NLO scale
dependence of $\pm 13\%$ corresponds to the $\pm 4.5\%$ variation
in $|V_{td}|$ shown in Table \ref{tab:vtdkpn}.)
\item
The NLO result is within the range estimated from scale dependence
at LO.
\item
The difference between the LO and the NLO result is about
$10\%$, hence a very moderate correction.
\item
The LO terms have the form $x\ln x (\alpha_s\ln x)^n$, resummed
to all orders $n$, and the NLO corrections are $x (\alpha_s\ln x)^n$.
(Here $x\equiv x_c$; due to the smallness of $x$ only first-order
terms need to be retained.)
At ${\cal O}(\alpha_s)$ ($n=1$) these terms are of the order
$\alpha_s x\ln^2 x$ and $\alpha_s x\ln x$, respectively.
The (unresummed) term of order $\alpha_s x$ contributes only
at NNLO in the charm sector and is {\it not\/} included in the
usual NLO results. However, this term is known from the full
${\cal O}(\alpha_s)$ calculation in the top sector and can be used
to estimate the truncation error, independently of the standard
procedure using residual scale dependence.
In fact, the ${\cal O}(\alpha_s x)$ term is about $10\%$ for charm,
fully compatible with the $\pm 13\%$ uncertainty estimated from
NLO scale dependence.
\end{itemize}
These observations demonstrate that perturbation theory
is well behaved for the charm contribution and that the
error estimate is under control.

So far we have considered the uncertainty due to truncation
of the resummed perturbative series. In addition there are also
power corrections. One source are the long-distance contributions
related to up-quark loops. They are of order $\Lambda^2_{QCD}/m^2_c$
and were estimated to be below $5\%$ \cite{HLLW}. The second
type of power corrections comes from higher orders in the OPE,
in the process where the charm quark is integrated out.
The leading corrections are of order $m^2_K/m^2_c$. They were recently
estimated to be again at the level of about $5\%$ \cite{FLP}.
These effects are safely below the perturbative uncertainty.

We conclude that the charm contribution to $K^+\to\pi^+\nu\bar\nu$
can be reliably computed based on OPE and RG improved perturbation
theory, and that the uncertainty can be assessed with confidence.

\subsection{Further Phenomenological Applications}

\begin{figure}[t]
  \vspace{6.5cm}
  \includegraphics{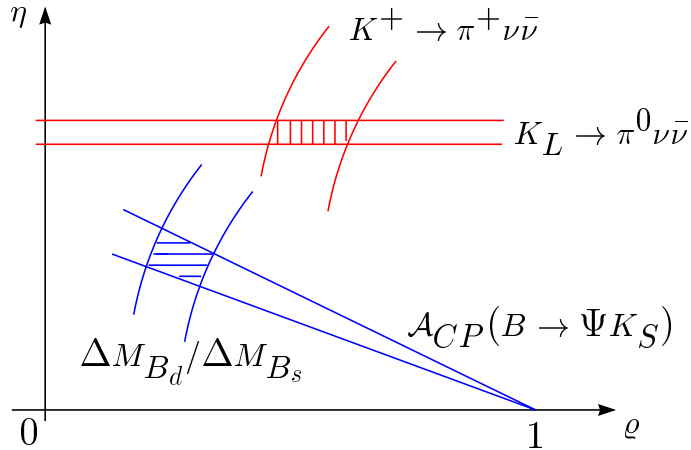}
  \caption{\it Schematic determination of the unitarity triangle
      vertex ($\varrho$, $\eta$) from $K\to\pi\nu\bar\nu$
      (vertically hatched) and from the $B$ system (horizontally
      hatched). Both determinations can be performed with small
      theoretical uncertainty and any discrepancy between them
      would indicate new physics, as illustrated in this 
      hypothetical example. 
    \label{fig:utkpn} }
\end{figure}

With a measurement of $B(K^+\to\pi^+\nu\bar\nu)$ and 
$B(K_L\to\pi^0\nu\bar\nu)$ available very interesting phenomenological
studies could be performed. 
For instance, $B(K^+\to\pi^+\nu\bar\nu)$ and 
$B(K_L\to\pi^0\nu\bar\nu)$ together determine the unitarity triangle
(Wolfenstein parameters $\varrho$ and $\eta$) 
completely (Fig. \ref{fig:utkpn}). 
The expected accuracy with $\pm 10\%$ branching ratio measurements is
comparable to the one that can be achieved by CP violation studies
at $B$ factories before the LHC era \cite{BB6}.

The quantity $B(K_L\to\pi^0\nu\bar\nu)$ by itself offers probably the
best precision in determining $\mbox{Im} V^*_{ts}V_{td}$ or,
equivalently, the Jarlskog parameter
\begin{equation}\label{jcp}
J_{CP}=\mbox{Im}(V^*_{ts}V_{td}V_{us}V^*_{ud})=
\lambda\left(1-\frac{\lambda^2}{2}\right)\mbox{Im}\lambda_t
\end{equation}
The prospects here are even better than for $B$ physics at the LHC.
As an example, let us assume the following results will be
available from $B$ physics experiments
\begin{equation}\label{lhcb}
\sin 2\alpha=0.40\pm 0.04\quad \sin 2\beta=0.70\pm 0.02\quad
V_{cb}=0.040\pm 0.002
\end{equation}
The small errors quoted for $\sin 2\alpha$ and $\sin 2\beta$ from
CP violation in $B$ decays require precision measurements at the LHC.
In the case of $\sin 2\alpha$ we have to assume in addition that the
theoretical problem of `penguin-contamination' can be resolved.
These results would then imply
$\mbox{Im}\lambda_t=(1.37\pm 0.14)\cdot 10^{-4}$.
On the other hand, a $\pm 10\%$ measurement 
$B(K_L\to\pi^0\nu\bar\nu)=(3.0\pm 0.3)\cdot 10^{-11}$ together with
$m_t(m_t)=(170\pm 3) GeV$ would give
$\mbox{Im}\lambda_t=(1.37\pm 0.07)\cdot 10^{-4}$. If we are optimistic
and take $B(K_L\to\pi^0\nu\bar\nu)=(3.0\pm 0.15)\cdot 10^{-11}$,
$m_t(m_t)=(170\pm 1) GeV$, we get
$\mbox{Im}\lambda_t=(1.37\pm 0.04)\cdot 10^{-4}$, a remarkable
accuracy. The prospects for precision tests of the standard model
flavour sector will be correspondingly good.

The future experimental prospects for $K^+\to\pi^+\nu\bar\nu$
and $K_L\to\pi^0\nu\bar\nu$ are discussed in the talks
by Bryman, Cox, Inagaki, Muramatsu and Ramberg.

Recent work on new-physics effects in $K\to\pi\nu\bar\nu$
can be found in \cite{NPH}.

\section{Summary}

Decays of kaons have played a key role in the development
of the standard model. Currently, flavour physics is entering
a new era of intense and promising experimental investigation. 
In this context rare $K$ decays in particular will continue to 
provide excellent opportunities. 

The search for decay modes forbidden in the standard model
($K_L\to\mu e$, $K\to\pi\mu e$) probes physics at very short
distances and possible exotic scenarios with impressive
sensitivity. Tests of chiral perturbation theory, beyond their
intrinsic interest, help to develop our theoretical understanding
of long-distance background to new physics effects, within
a model-independent approach to low-energy QCD. Processes of interest 
here are $K^+\to\pi^+ l^+l^-$, $K_L\to\pi^0\gamma\gamma$, 
$K_S\to\gamma\gamma$ among many others.
Specific further opportunities are given by searching for
violations of discrete symmetries (CP, T) in $K_L\to\pi^0 e^+e^-$
or $K^+\to\pi^0\mu^+\nu$ ($\mu$-polarization).
Of particular importance are standard model precision tests
with the golden modes $K^+\to\pi^+\nu\bar\nu$ and $K_L\to\pi^0\nu\bar\nu$.

The main goals of flavour physics will be accurate and decisive
tests of the CKM mechanism and the search for new phenomena.
In this respect kaon physics can contribute unique information,
complementary to physics with $B$ and $D$ mesons. At present, the
standard model appears to work well, also in the flavour sector,
and has passed already a number of nontrivial tests.
Therefore theoretically clean, high-precision observables,
such as those offered by rare $K$ decays, will become even more
valuable and important.

\end{document}